\documentstyle[12pt]{article}
\title{INTERACTION OF  WILSON LOOPS IN THE  SU(N) GAUGE
THEORY}
\vspace{2cm}
\author{ A.Yu.Dubin and Yu.S.Kalashnikova \\ Institute of Theoretical and
Experimental Physics\\ 117259,
Moscow}
\date{}
\newcommand{\be}{\begin{equation}}
\newcommand{\ee}{\end{equation}}
\begin{document}
\maketitle

\begin{abstract}

The average of two Wilson loops is expressed in terms of gauge invariant
field strength correlators. Assuming the existence of finite
 correlation  length $T_g$ and taking into account the
 absence of a fixed direction in colour space, we generalize the area law
 asymptotics for the case of  the average of two Wilson loops embedded into
 the same plane.   The results  are  presented in terms of
 averaged  single Wilson loop operators in irreducible representations which
 correspond to the  geometry of the contours. A  special reduction of our
 results (obtained for arbitrary dimensionality of space-time) to the case
 of two-dimensional SU(N) theory is performed and the connection to
 the results existing in literature is  established.
   \end{abstract}

   \section{Introduction}

   To analyse the hadronic
 spectra and interactions one is to consider the Green functions for many
 quark system in the confining gluonic fields.  Making use of the
 Feinman--Schwinger proper time representation [1], these Green functions can
 be expressed in terms of the average of Wilson loops corresponding to the
 hadrons participating in the process under consideration. In the simplest
 case of the $q\bar{q}$ system the area law for $<W(C)>$ enables one to
 study the dynamics of quarks connected by the string. The quantization of
 the "minimal" QCD string (corresponding to the minimal area asymptotics for
 $<W(C)>$) with quarks has been done in [2].  For high energy scattering the
 interaction of two Wilson loops was considered in [3] in eikonal type
 approach, while for low and intermediate energies the problem of evaluating
 decay and scattering amplitudes in terms of Wilson loops average is still
 awaiting for the solution.

  The purpose of the present paper is to  derive the expressions for
  the average  of two Wilson loops in terms of gauge invariant field
   strength correlators, and to obtain the generalization of the
   area law  asymptotics for this case.

         We use the cluster expansion method [4,5] which enables one to
   write out the Wilson loops average in terms of irreducible in space
   correlators (the so--called cumulants). Our basic assumption is the
   existence of the field strength cumulants with  finite
   correlation length $T_g$ [5,6], which have the structure of
   Lorentz indexes containing the part expressed only in terms of
   Kronecker delta functions $\delta_{\mu\nu}$. The second order
   correlator of this type (defined in the gauge invariant way [5])
   $$\ll F_{\mu\nu} (x) F_{\rho\sigma}(y)\gg\sim
   (\delta_{\mu\rho}\delta_{\nu\sigma}-\delta_{\mu\sigma}\delta_{\nu\rho})$$
   was measured recently by lattice simulations [7].  We also make extensive
   use of the condition that there is no fixed direction in colour space.
   In the case of one loop the presence of such cumulants leads to
   the appearence of long range interaction between quarks and
   allows to obtain the minimal area law asymptotics for
   large contours [5], which is also observed on the lattice.

    We present the generalization of the area law for the case when our
   loops are embedded into the same plane (Fig.1), or into the planes
   with distance between them less than the gluonic correlation length
   $T_g$.  To obtain the generalized area law it appears necessary to reduce
   the initial SU(N) matrix problem to the effective scalar one with  short
   range scalar cumulants. While  a single Wilson loop corresponds to the
   circulation of colour charge in fundamental representation along the
   contour, for the case of two loops (with the geometry of Fig.1) one
   actually has the circulation of colour charges in various irreducible
   representations in accordance with the decomposition $$
N\bigotimes\bar{N}= singlet ~\bigoplus~adjoint $$  for oppositely directed
 loops, and $$      N\bigotimes {N}= symmetric ~\bigoplus~antisymmetric $$
 for parallelly directed loops,so that the average of two loops can be
  (symbolically) presented as
    $$
   <W(C_1)^{\alpha}_{\beta}W(C_2)^{\gamma}_{\delta}>= $$ $$(singlet
   )^{\alpha}_{\delta}

(singlet)^{\gamma}_{\beta}+(N^2-1)(adjoint)^{\alpha}_{\delta}(adjoint)^{\gamma}_{\beta}
   $$
   or
   $$
   <W(C_1)^{\alpha}_{\beta}W(C_2)^{\gamma}_{\delta}>=$$$$
      \frac{N(N+1)}{2} (sym)^{\alpha\gamma }(sym)_{\beta\delta} +
      \frac{N(N-1)}{2}(antisym)
   ^{\alpha
   \gamma}
   (antisym)_{\beta\delta}
  $$

   As a consequence, for large contours in quasiflat geometry the following
   expressions are  obtained:
     $$<SpW(C_1)SpW(C_2)>= \sum_{r} D^{(r)}
   exp\{-\sigma^{(f)}(S_1-S_2)-\sigma^{(r)}S_2\},~~$$
   $$~~~~~~~~~~~~~~~~~~~~~~~~~~~~~~~~~~~~~~~S_1>S_2;~~S_1,S_2\gg T_g^2$$
   where $D^{(r)}$ is the dimension of the given irreducible representation
   $\{r\},~~\sigma^{(r)}$ is the corresponding string tension, and
   $\sigma^{(f)}$ is the fundamental string tension entering the area law
   for a single loop. Therefore we connect the average of two Wilson loops
   (in fundamental representation) to the average of single Wilson loop in
   representations in accordance with the geometry of the contours. The
   expressions for the selfintersecting loop (with the geometries of Fig.2)
   are also obtained.

     We perform the  reduction of our resuts to the 1+1
   gauge theory, and find the agreement with the calculations [8,9] based on
   the Migdal--Makeenko equations [10] and non-abelian Stokes theorem [11].
   We note that recently in 1+1 QCD  the  average of arbitrary number
   of Wilson loops on  an arbitrary two-dimensional
   manifold has been calculated and  the relation  to the
   string theory of maps  was obtained [12].

 Correlation of Wilson loops was also studied numerically in
 lattice simulations [13]. The case when one of  the loops is small and
 serves as a probe to study the field distribution around the string is a
 topic of a vivid interest for several years  [14-16]. When the loops
 are large, both measurements and interpretation are much more difficult and
 accuracy is not yet enough.

 The plan of the paper is as follows.

 In Section 2 we introduce  the cluster expansion on the example of a single
 Wilson loop, using the Fock--Schwinger gauge to simplify our
 derivation.

 In Section 3 we derive expressions in terms of the cumulants
  for the average of two loops.

 The explicit answers in terms of the corresponding string tensions and areas
 are given in Section 4 for the case of large contours embedded
 into the same plane.

 Discussion and comparison with the two-dimensional
 $SU(N)$ theory is done in Section 5.

 Two appendices A and B are devoted to the derivation of scalar
 clusterization condition and the relations between string
 tensions for 1+1 SU(N) theory  respectively.

 \section{Cluster expansion method.}

 Before proceeding with the case of two loops we remind the essentials
 concerning one loop; the details may be found in [5]. The quantity to be
 considered here is
 \be
 <W(C)> = <Sp~P~exp~ig \int_CA_{\mu}dx_{\mu}>,
 \ee
 where brackets mean the average over all gluonic fields with the QCD
 weight. The expression (1) may be rewritten using the non-abelian Stokes
 theorem [5,11,17] in the explicitly gauge--invariant form
 \be
 <W(C)> = <Sp~P~exp~ig \int_S d\sigma_{\mu \nu}(x)F_{\mu\nu}(x,x_0)>,
 \ee
 where $S$ is the minimal surface bounded by the contour $C$, and
 \be
 F_{\mu\nu}(x,x_0)= \Phi(x_0,x)F_{\mu\nu}(x)\Phi(x,x_0),
 \ee
 $\Phi(x_0,x)=P~exp~ig \int^x_{x_0}A_{\mu}dz_{\mu}$ is the parallel
 transporter along the path connecting the points $x$ and $x_0$.

 We shall use at the intermediate stages the Fock--Schwinger gauge
 \be
 A^a_{\mu}(x)(x-x_0)_{\mu}=0,
 \ee
 $$A^a_{\mu}(x)=\int^1_0\alpha
 d\alpha(x-x_0)_{\nu}F^a_{\nu\mu}((x-x_0)\alpha+x_0)
 $$
  for which $\Phi(x_0,x)$ in (3) is equal to unity, and choose the point
 $x_0$ inside the contour, as it is shown at Fig.3;  for such choice the
 integration in (1) takes place actually along the identically deformed
 contour of Fig.3, and (2) follows from (1) in the most straightforward way.

 We present the expression (1) in the form
 \be
 <W(C)> = N\sum^{\infty}_{n=0}(ig)^nI_n~,~~I_0=1,
 \ee
 $$~~~~~~~~~~~I_n=\frac{1}{N} P\int dx_1...dx_n<Sp~(A(x_1)...A(x_n))>
\hfill~~~~~~~~~~~~~~~~~~~~~~ (5^a)$$
with ommited Lorentz indices.
 For the contours without selfintersections the coordinates $\{x_i\}$ on the
 contour $C$ can be parametrized in the unique way by ordering parameres
 $\{s_i\}, 0<s_i<1$, which are determined, in turn, by angular variables
 $\{\theta_i\}$ along the contour (see Fig.3). It is naturally to assume
 that for the ensemble of stochastic fields the angular  clusterization
 condition holds true, \be <(A(s_1)...A(s_m) )^{\alpha}_{\beta}
 (A(s_{m+1})...A(s_{m+n}) )^{\rho}_{\sigma}>\to
  \ee
   $$ \to<(A(s_1)...A(s_m)
 )^{\alpha}_{\beta} > <(A(s_{m+1})...A(s_{m+n}) )^{\rho}_{\sigma}> $$
 if the
 cluster of points $\{s_1...s_{m+n}\}$ is separated spatially into two
 clusters, $\{s_1...s_m\}$ and $\{s_{m+1}...s_{m+n}\}$, so  that
 $$|x_i-x_j|\gg T_g,~~i\epsilon\{1...m\},~~ j\epsilon\{m+1..m+n\},$$ where
 $T_g$ is the gluonic correlation length.

   Due to the condition (6) the correlators of a given order in  eq.(5)
 contain the smaller order correlators as disconnected parts (the parts
 which do not vanish at large intercluster  distances), and it is useful to
 represent $<W(C)>$ in terms of the connected correlators, the so--called
 cumulants [4]. It can be done by  calculation of $ln <W(C)>$. Formally,
 the cumulants ${\cal K}^{\alpha}_{m\beta}(x_1...x_m)$ are defined from the
 equation  [4]
  \be
 <(W(C))^{\alpha}_{\beta}>=\sum^{\infty}_{n=0}(ig)^n(I_n)^{\alpha}_{\beta}=
 \ee
 $$
 \{ exp \sum^{\infty}_{m=1}\frac{(ig)^m}{m!}{\cal K}_m\}^{\alpha}_{\beta},$$
 where
 \be
 ({\cal K}_m)^{\alpha}_{\beta}= \int
 dx_1...dx_mK_m^{\alpha}_{\beta}(x_1...x_m).  \ee

 It  was shown in  [4] that if the clusterization condition (6) takes place,
  then the functions
 $K_m^{\alpha}_{\beta}(\{x_m\})$ defined  in accordance with eqs.(7), (8)
 vanish if the cluster of points $\{x_m\}$ is  separated spatially into any
 number of clusters $\{x_{m_1}\},\{x_{m_2}\}...$ .  Equivalently,  for large
 smooth surfaces  the field
 strength cumulants (with the Lorentz indexes structure, which contains the
 part expressed only in terms of Kronecker delta functions) coming to eq.(8)
 from the expression (4) for Fock-Schwinger gauge lead to the following
 result for the corresponding integrals [5]
  \be \frac{(ig)^m}{m!} ({\cal
 K}_m)^{\alpha}_{\beta}=(\sigma_m)^{\alpha}_{\beta}S+(\mu_m)^{\alpha}_{\beta} p
 \ee
 where $S/p$ is the area/perimeter of the surface, and $\sigma,\mu$
 may be (for arbitrary stochastic ensemble) non-unity matrices. Large and
 smooth means that both the size of the contour and the averaged size of
 contour irregularities are much larger than the correlation length $T_g$.

 The equation (9), however, is not sufficient in general for the "scalar"
 area law
 $$<Sp~W(C)>\to exp(-\sigma S)$$
 we are to have in QCD. The latter is achieved if there is no fixed
 colour direction in the vacuum, i.e., in addition to eq.(6), the condition
 \be
 <({A}(s_1)...{A}(s_n))^{\alpha}_{\beta}>=
\delta^{\alpha}_{\beta}{\cal A}(s_1...s_n),
 \ee
 $${\cal A}(s_1...s_n)=\frac{1}{N}<Sp({A}(s_1)...{A}(s_n))>.$$
 is imposed.  Then the "matrix" area law (9) is obviously reduced to the
 "scalar" one,
 $$({\cal K}_m)^{\alpha}_{\beta}=\delta^{\alpha}_{\beta}{\cal
 K}_m,$$ $$(\sigma_m)^{\alpha}_{\beta}=\delta^{\alpha}_{\beta}\sigma_m~,~~
 (\mu_m)^{\alpha}_{\beta}=\delta^{\alpha}_{\beta}\mu_m,$$
 and for large smooth contours one has [5], up to shape corrections,
 \be
 <Sp~W(C)>=N exp(-\sigma S-\mu p),
 \ee
 where
 \be
 \sigma=\sum^{\infty}_{m=2}\sigma_m~,~~
 \mu=\sum^{\infty}_{m=2}\mu_m~.
 \ee
 (It is supposed that  seria (12) for $\sigma$ and $\mu$ converge).

   The conditions (6)
  and (10) can be combined into the "scalar" clusterization condition
  for the functions ${\cal A}$
  \be
  \frac{1}{N}<Sp(A(s_1)...A(s_{m+n}))>\to \ee $$ \to
  \frac{1}{N}<Sp(A(s_1)...A(s_{m}))>\cdot
  \frac{1}{N}<Sp(A(s_{m+1})...A(s_{m+n}))>$$
     if the cluster $\{s_1...s_{m+n}\}$ is separated into two clusters
 $\{s_1...s_{m}\}$  and\\  $\{s_{m+1}...s_{m+n}\}$.

   This condition enables
 one to reduce the matrix exponentiation (7) to the scalar one which will
 be necessary for two loops case analysis.  To this end we are to tranform
 the ordered in $\{s\}$ integration $ s_{i+1}>s_i$, into the
 integration over the whole region $0<s_j<1$, so that   $I_n$ in (5) can
 be rewritten as \be I_n=P\int ds_1...ds_n{\cal A}(s_1...s_n)= \ee
 $$~~~~~~~~~~~~~~~~~~~~~~~~\frac{1}{n!}\int(ds_i)^n{\cal
 A}(s_1...s_n)~~~~~~~~~~~~~~~~~~~~~~~~~~~~~~~~~(14^a)$$
  where the ordering
 of colour  matrices in ${\cal A}(s_1...s_{n})$  in $ (14^a)$ follows the
 ordering of variables $\{s_i\}$.
 In such representation the application of ordinary scalar exponentiation
 [4] to eq. (5)  gives the definition of scalar cumulants $K_m$, vanishing
 at large distances due to eq. (13):  $$ \frac{1}{2!}K_2(x_1x_2)={\cal
  A}(x_1x_2)$$ \be \frac{1}{3!}K_3(x_1x_2x_3)={\cal A}(x_1x_2x_3) \ee
 $$\frac{1}{4!}K_4(x_1x_2)={\cal
 A}(x_1x_2x_3x_4)-{\cal A}(x_1x_2){\cal A}(x_3x_4)-{\cal A}(x_1x_3)
 {\cal A}(x_2x_4)-{\cal A}(x_1x_4){\cal A}(x_2x_3)$$
 and so on. Therefore, we demonstrate that the condition (13) leads to
 scalar area law (11).

 \section{Two Wilson loops average}

         We start with the evaluation of
         \be
         <(W(C_1))^{\alpha}_{\beta}(W(C_2))^{\gamma}_{\delta}>
  \ee
  for two contours in the same plane inserted one into
  another. We have deformed the contours identically by making tentacles
  from the open ends of the contours to the point $x_0$ inside the both
  loops (Fig.4). This point participates in the definition (3) of the
  $F_{\mu\nu}(x,x_0)$ and in the Fock--Schwinger gauge (4), which is used
  in what follows. The straightforward application of the exponentiation
  procedure (7) to the expression (16) leads only to the "matrix" area law,
  as it is clear already from the analysis of the one loop case. Moreover,
  in contrast to the one loop case, the condition (10) does not lead, as we
  shall see, automatically to the area law.

  To develop a regular method we are to express the average (16) in terms of
  the proper scalar short range correlators. Colour structure of the
  expression (16) which can be interpreted as the circulation of two
  fundamental colour charges, is decomposed into the pairs
  corresponding to the circulation of charges in  irreducible
  representation  accordingly to the formulae
  \be
  N\otimes\bar{N}= singlet \oplus adjoint
  \ee
   for the oppositely directed
  loops, and \be N\otimes N = symmetric \oplus antisymmetric \ee for the
  parallelly directed loops.

  We prove that scalar correlators corresponding to each term of the
  decomposition of $<SpW(C_1)SpW(C_2)>$ in accordance with (17),(18)
  satisfy the scalar clusterization condition similar to (13).
  Consequently, the application of scalar exponentiation  separately to each
  part  of this decomposition gives the
  definition of short range  scalar cumulants and  generalized area
  law.

          Instead of  eq. (6) for two loop case we consider the average of the
type
          \be
  <(A(x_1^{(1)})...A(x_m^{(1)}))^{\alpha}_{\beta}
  (A(x_1^{2})...A(x_n^{(2)}))^{\gamma}_{\delta}>
  \ee
  where points $\{x_i^{(1)}\}$ belong to the contour $C_1$ and
  points $\{x^{(2)}_k\}$ belong to the contour $C_2$. First we
  generalize the clusterization condition (6) for our two loops case,
  and prove that it leads with necessity to the decomposition (17), (18)
   for the definition of short range scalar cumulants.

     Due to the
  interaction between the loops the average (19) is not reduced to the
product of averages if the points on both contours are correlated in angles
defined with respect to a halfline, starting at $x_0$.  Therefore  only if
the cluster of points $\{s_1^{(1)}...s_{m+n}^{(1)},
s_1^{(2)}...s_{k+l}^{(2)}\}$ is separated in angles into two clusters
$\{s_1^{(1)}...s_{m}^{(1)}, s_1^{(2)}...s_{k}^{(2)}\}$ and
$\{s_{m+1}^{(1)}...s_{m+n}^{(1)}, s_{k+1}^{(2)}...s_{k+l}^{(2)}\}$ (the
superscripts (1) and (2) stand to label the points from the first and the
second loop), the generalized clusterization condition holds true for the
ensemble of stochastic fields:  \be
<(A(s_1^{(1)})...A(s^{(1)}_m)^{\alpha}_{\beta}
(A(s_{m+1}^{(1)})...A(s^{(1)}_{m+n})^{\gamma}_{\delta }\cdot
\ee
$$\cdot (A(s_{k+l}^{(2)})...A(s^{(2)}_{k+1}))^{\rho}_{\sigma}
(A(s_{k}^{(2)})...A(s^{(2)}_{1}))^{\mu}_{\nu}>\to$$
$$\to <(A(s_{1}^{(1)})...A(s^{(1)}_{m}))^{\alpha }_{\beta}
 \cdot (A(s_{k}^{(2)})...A(s^{(2)}_{1}))^{\mu}_{\nu}>\cdot
$$
$$\cdot <(A(s_{m+1}^{(1)})...A(s^{(1)}_{m+n})^{\gamma}_{\delta } \cdot
(A(s_{k+l}^{(2)})...A(s^{(2)}_{k+1}))^{\rho}_{\sigma}>$$
for opposite ordering of loops, and
$$~~~~~~~~~~~~~~~~<(A(s_1^{(1)})...A(s^{(1)}_m)^{\alpha}_{\beta}
(A(s_{m+1}^{(1)})...A(s^{(1)}_{m+n})^{\gamma}_{\delta
}>\cdot~~~~~~~~~~~~~~~~~~~~(20') $$
$$\cdot(A(s_{1}^{(2)})...A(s^{(2)}_{k}))^{\mu}_{\nu}
(A(s_{k+1}^{(2)})...A(s^{(2)}_{k+l}))^{\rho}_{\delta}>\to
$$
$$\to <(A(s_{1}^{(1)})...A(s^{(1)}_{m}))^{\alpha }_{\beta}
  (A(s_{1}^{(2)})...A(s^{(2)}_{k}))^{\mu}_{\nu}>\cdot
$$
$$\cdot <(A(s_{m+1}^{(1)})...A(s^{(1)}_{m+n})^{\gamma}_{\delta } \cdot
(A(s_{k+1}^{(2)})...A(s^{(2)}_{k+l}))^{\rho}_{\sigma}>$$
for parallel ordering.
 Again, as for the case of one
  loop, in the Fock-- Schwinger gauge with the point $x_0$
  inside the loops the clusterization condition (20) is actually the
  angular one.

  In what follows the shorthand notations are used:
  \be
  <(A(s^{(1)}_{1})...A(s^{(1)}_{i})^{\alpha}_{\beta} (A(
  s^{(2)}_{j})...A(s^{(2)}_{1}))^{\gamma}_{\delta}>=
  <(a_i)^{\alpha}_{\beta} (b_j)^{\gamma}_{\delta}>\ee
and
$$
{}~~~<(A(s^{(1)}_{1})...A(s^{(1)}_{i})^{\alpha}_{\beta} (A(
  s^{(2)}_{1})...A(s^{(2)}_{j}))^{\gamma}_{\delta}>=
  <(a'_i)^{\alpha}_{\beta} (b'_j)^{\gamma}_{\delta}>.~~~~~~~~~(21')$$

 To reduce, as in one loop case, the  SU(N) matrix problem to the effective
      scalar one we first note, that the natural generalization of eq.(10)
 is the following expression for the four--point averages $(21)$ and $(21')$
 in terms of two scalar functions ${\cal A,B}$

   \be <(a_i)^{\alpha}_{\beta}(b_j)^{\gamma}_{\delta}>= {\cal
A}(i,j)\delta^{\alpha}_{\beta}\delta^{\gamma}_{\delta}+ {\cal
B}(i,j)\delta^{\alpha}_{\delta}\delta^{\gamma}_{\beta} \ee \be
      <(a'_i)^{\alpha}_{\beta}(b'_j)^{\gamma}_{\delta}>=
{\cal A}'(i,j)\delta^{\alpha}_{\beta}\delta^{\gamma}_{\delta}+
{\cal B}'(i,j)\delta^{\alpha}_{\delta}\delta^{\gamma}_{\beta}
 \ee
      instead of one function ${\cal A}$ of eq.(10).
It is demonstrated in Appendix A that the  use of the decomposition (17)
      leads to the following scalar correlators, satisfying the scalar
clusterization condition \be
 \frac{1}{N}<Sp(a_{i1+i2}b_{j2+j1})>\to \ee
$$\to \frac{1}{N}<Sp(a_{i1}b_{j1})>\cdot \frac{1}{N}<Sp(a_{i2}b_{j2})>$$ and
\be \frac{2}{N^2-1}<Sp(a_{i1+i2}\lambda_a b_{j2+j1} \lambda_a)>\to \ee $$\to
\frac{2}{N^2-1}<Sp(a_{i1} \lambda_b b_{j1} \lambda_b )>\cdot
\frac{2}{N^2-1}<Sp(a_{i2}\lambda_c b_{j2} \lambda_c)>$$
with
\be
\frac{1}{N}<Sp(a_{i}b_{j})>=
{\cal A}(i,j)+N{\cal B}(i,j)= J^{(0)}_{ij}
 \ee
 and
 \be
\frac{2}{N^2-1}<Sp(a_{i}\lambda_a b_{j} \lambda_a)>= A(i,j)=
J^{(adj)}_{ij}.
\ee
Similarly, from the decomposition (18) we have
\be
\frac{1}{N(N\pm 1)}<\{Sp(a'_{i1+i2}) Sp(b'_{j1+j2})\pm
Sp(a'_{i1+i2}b'_{j1+j2})\}>\to
\ee
$$\frac{1}{N(N\pm 1)}<\{Sp(a'_{i1}) Sp(b'_{j1})\pm
Sp(a'_{i1}b'_{j1})\}>\cdot $$
$$\cdot \frac{1}{N(N\pm 1)}<\{Sp(a'_{i2}) Sp(b'_{j2})\pm
Sp(a'_{i2}b'_{j2})\}> $$
with
\be
\frac{1}{N(N\pm 1)}<\{Sp(a'_{i}) Sp(b'_{j})\pm
Sp(a'_{i}b'_{j})\}>=
\ee
$$={\cal A}'(i,j)\pm{\cal B}'(i,j)=J_{ij}^{(s,a)}$$

To take advantage of these scalar correlators we decompose the pairs of
colour indeces in  $<W(C_1)^{\alpha}_{\beta}W(C_2)^{\gamma}_{\delta}>$ in
accordance with the decomposition (17),(18), that corresponds to the
following:
 \be <SpW(C_1)SpW(C_2)>=
\ee $$ \frac{1}{N}< Sp(W(C_1)W(C_2))>+
2<Sp(W(C_1)\lambda_aW(C_2)\lambda_a>$$
 with
 \be
<Sp(W(C_1)W(C_2)>=N\sum_n(ig)^n~I_n^{(0)} \ee \be
<Sp(W(C_1)\lambda_aW(C_2)\lambda_a>=\frac{N^2-1}{2}\sum_n(ig)^nI_n^{(adj)},
\ee
for the case of opposite direction in the loops, and to
\be
<Sp(W(C_1)) Sp(W(C_2))>=
\ee
$$
\frac{1}{2}\{ <SpW(C_1) SpW(C_2)+
<Sp(W(C_1) W(C_2))>\}+$$
$$+\frac{1}{2}\{
<SpW(C_1)SpW(C_2)>-<Sp(W(C_1) W(C_2))>\}$$
with
\be
<SpW(C_1)SpW(C_2)\pm SpW(C_1) W(C_2))>=(N\pm 1)\sum_n(ig)^nI_n^{(s,a)}
\ee
for the case of  parallel directions. The quantities $I_n^{(r)}$ ($(r)$ stand
for singlet (0), adjoint  (adj), symmetric (s) and antisymmetric (a)
representations) are
$$I_n^{(r)}=\sum_{i+j=n}I_{n,ij}^{(r)}~,~~I^{(r)}_0=1~,$$
\be
I_{n,ij}^{(r)}=P\int
ds_1^{(1)}...ds_i^{(1)}ds_1^{(2)}...ds_j^{(2)}J^{(r)}_n(i,j)
\ee
and $J_n^{(r)}(i,j)$  are defined in (26),(27) and (29).

In order to perform  the final reduction to the scalar case, as it was
done for one loop (see eq.(14$^a$)), one is to represent the
integration in (35) over the ordered region as the integration over the
whole region:
\be
I_n^{(r)}= \sum_{i+j=n}I_{n,ij}^{(r)} =
\sum_{i+j=n} \frac{1}{i!}\frac{1}{j!}\int(ds^{(1)})^i(ds^{(2)})^j
J^{(r)}_n(i,j)=
\ee
$$
=\sum_{i+j=n}\frac{1}{n!}\int(ds)^{i+j}J^{(r)}_n(i+j)
$$
 where the integration is
performed over the whole contour $C_{1}+C_2$, and the ordering of the fields
(and $\lambda$ matrices -- insertions  in eq. (27)) follows the ordering in
$\{s\}$.Due to the eqs. (24)-(29) the averages $J_n^{(r)}(i+j)$ satisfy
the clusterization condition for angularly separated clusters. Therefore the
scalar exponentiation procedure applied separately to  each term in
eqs.(30),(33) leads to the cumulants which vanish  [4] if the  points
on loops from any clusters are  separated in angles.

The result of  such exponentiation is
\be
<Sp(W(C_1)SpW(C_2) )>=
\ee
$$exp \sum^{\infty}_{m=2}\frac{(ig)^m}{m!}{\cal K}^{(0)}_m+
+(N^2-1)exp \sum^{\infty}_{m=2}\frac{(ig)^m}{m!}{\cal K}_m^{(adj)}$$
for oppositely directed loops, and
\be
<Sp(W(C_1)Sp(W(C_2) )>=
\ee
$$\frac{N(N+1)}{2} exp
\sum^{\infty}_{m=2}\frac{(ig)^m}{m!}{\cal  K}^{(s)}_m+
\frac{N(N-1)}{2}exp
\sum^{\infty}_{m=2}\frac{(ig)^m}{m!}{\cal K}_m^{(a)}$$
We have introduced  the integrated cumulants ${\cal K}_m^{(r)}$
\be
{\cal K}_m^{(r)}=\sum_{i+j=m}{\cal K}^{(r)}_{m,ij}
\ee
\be
{\cal K}^{(r)}_{m,ij}=\int
dx_1^{(1)}...dx_i^{(1)}dx_1^{(2)}...dx_j^{(2)}K_{m,ij}^{(r)}(x_1^{(1)}...x_i^{(1)}x_1^{(2)}...x_j^{(2)})
\ee
which are related to  $J_m^{(r)}$ of eq. (36) by  the recurrent
procedure similar to (7),(15).

\section{ Area law for the average of two Wilson loops}

Let's analyse the expressions (37),(38) for the large contours
$C_1$ and $C_2$ of Fig.4, and obtain the generalization of the area law
(using the Fock--Schwinger gauge (4)).  In what follows we neglect the
 perimeter type contributions, which always come as in one loop case (11).
 The integrals (40) contain, in addition to "angular" integration over
$\{s_i\}$, the integration over the radial variables $(xd\alpha)_{i} $,
see eq.(4); this double integration can be rewritten as the summation over
the corresponding sectors. Due to the existence of short range cumulants for
a given angle only the overlaps of the sectors beloning to different loops
contribute to the area law (see Fig.3).  Therefore, only the ${\cal
K}^{(r)}_{m,m0}$ are proportional (in the large area limit) to
$S_1(S_1>S_2)$, and the ${\cal K}^{(r)}_{m,ij},$ with $j\not=0$ are
proportional to $S_2$.  We stress again that we assume that the cumulants
$\ll F_{\mu\nu}... F_{\rho\sigma}\gg$ contain the part expressed only in
terms of Kronecker delta functions (see [5] for the details). It is this
part which originates the long range interaction between quarks leading to
the area law.

First we note that due to the condition (10) we have for  ${\cal
K}_{m,mo}^{(r)}(S_1>S_2)$
\be
 \frac{(ig)^m}{m!}{\cal K}_{m,m0}^{(r)}=\sigma_m^{(f)}S_1,\ee
where  $\sigma_m^{(f)}=\sigma_m$ is the contribution to the fundamental
string tension of eq.(12) which arises from the cumulant of m-th order.
 Second, due to the finite gluonic correlation length all the ${\cal
K}^{(r)}_{m,ij}$ with $ j\not=0$ are proportional to $S_2$, and therefore
one can reduce the integration over $C_1+C_2$ to the integration over
$C'_2+C_2,$ where  $C'_2$ coincides with
$C_2$ and directed as $C_1$.  Consequently,
$$\frac{(ig)^m}{m!}\sum_{i+j=m,j\not{=}0}{\cal
K}^{(r)}_{m,ij}=-\sigma^{(f)}S_2+ \frac{(ig)^m}{m!}\sum_{i+j=m}\tilde{{\cal
K}}^{(r)}_{m,ij},$$ where $\tilde{{\cal K}}^{(r)}_{m,ij}$ is defined by
integral (40) over $C'_2+C_2$.

 Observing that \be
 D^{(r)} exp
(\sum_m\frac{(ig)^m}{m!}\sum_{i+j=m}\tilde{{\cal K}}^{(r)}_{m,ij})=<Sp~
P~exp~ig \int_{C_2}(A_{\mu}^a\Lambda^{(r)}_a)dx_{\mu}>\ee
 where
$\Lambda^{(r)}$ and $ D^{(r)}$ are the generators and dimensions of the
corresponding irreducible representations $r$, we finally get   the
following expression: \be
\sum_m\frac{(ig)^m}{m!}{\cal K}_m^{(r)}=
\sigma^{(f)}(S_1-S_2)+\sigma^{(r)}S_2
\ee
 (The string tension $\sigma^{(0)}$ in the
singlet representation is obviously equal to zero).

 Now we arrive to the main result of the paper.
 The area law  that follows from the equations (37), (38) reads
 \be
 <Sp(W(C_1))Sp(W(C_2))> = exp (-\sigma^{(f)}(S_1-S_2)) +
 \ee
 $$+(N^2-1) exp (-\sigma^{(f)}(S_1-S_2) - \sigma^{(adj)}S_2),{S_1>S_2}$$
 for oppositely directed loops, and
 \be
 <Sp~(W(C_1))Sp~(W(C_2))> =\frac{N(N-1)}{2} exp (-\sigma^{(f)}(S_1-S_2))
 -\sigma^{(a)}S_2)+
 \ee
 $$
 +\frac{N(N+1)}{2} exp (-\sigma^{(f)}(S_1-S_2))
 -\sigma^{(s)}S_2)$$
 for parallel loops, with $\sigma^{(adj)},~~\sigma^{(a)}$ and $\sigma^{(s)}$
 and $\sigma^{(s)}$ given by (43). We stress here that due to eq. (42), (43)
 we express the average of two Wilson
 loops (in fundamental representation) in terms of the proper combinations
 of single Wilson loop averages in representations corresponding to the
 geometry of the contours.

 If the main contribution to the string tension arises
 from the lowest cumulant $K_2$, then, as it is shown in the Appendix B,
the string tensions $\sigma^{(r)}$ are related to each other by the formulae
 \be\sigma^{(adj)}=\frac{2N^2}{N^2-1}\sigma^{(f)}~,
 \ee

$$~~~~~~~~~~~~~~~~~~~~~~~~~~ \sigma^{(a)}=\frac{2(N-2)}{N-1}\sigma^{(f)}~,
 ~~~~~~~~~~~~~~~~~~~~~~~~~~(46')$$
  $$
 ~~~~~~~~~~~~~~~~~~~~~~~~~~~~\sigma^{(s)}=\frac{2(N+2)}{N+1}\sigma^{(f)}~.
 ~~~~~~~~~~~~~~~~~~~~~~~~~~~(46^{\prime\prime})$$
that follow from the relations between quadratic Casimir operators.
 Lattice calculations [18] demonstrate
 that the relation (46) between $\sigma^{(f)} $ and $\sigma^{(adj)}$ holds
 with a  good accuracy. This fact allows to hope that the series (12) for
 string tension is saturated to a large extent by the second order
 contribution.

One can also write out the expressions for the average of a loop with
 selfintersections. For a contour with one selfintersection the usual area
 law
 $$ <Sp(W(C_1))W(C_2)>= N exp
 (-\sigma(S_1+ S_2)) $$
  takes place, as it is clear from the given above considerations, only if
 the areas inside both loops do not overlap (Fig.4$^a$). In the case of
 Fig $4^b$ the average can be calculated similarly to the case (33) of
   parallelly directed loops:  $$ <Sp(W(C_1)W(C_2))>= \frac{1}{2}\{
 <SpW(C_1)SpW(C_2)>+<Sp(W(C_1)W(C_2))>\}- $$ $$ - \frac{1}{2}\{
 <SpW(C_1)SpW(C_2)>-<Sp(W(C_1)W(C_2))>\}$$ with the result
 \be
 <Sp(W(C_1)W(C_2))>=
\frac{N(N+1)}{2} exp
 (-\sigma^{(f)}(S_1-S_2)-\sigma^{(s)}S_2)-
 \ee
  $$- \frac{N(N-1)}{2}
 exp (-\sigma^{(f)}(S_1-S_2)-\sigma^{(a)
 }S_2).$$

 We note that eq.(47) is less then zero when

$$ \frac{N+1}{N-1}< exp (+(\sigma^{(s)}-\sigma^{(a)})S_2)$$
 which is achieved for $\sigma^{(r)}$ determined by eqs (46) if $S_2$ is
 sufficiently large.

   Let us briefly discuss what result is to be expected if
 the contours are not in the same plane. If the average distance between the
 contours is less than the correlation length $T_g$, in the limit of large
 areas one obtains the same expressions (44)-(45). If the distance between
 the planes is larger then $T_g$, the interaction between the loops, in
 contrast to  the flat geometry case, becomes small. To extract the factor
 corresponding to interaction we first gather the terms ${\cal
 K}^{(r)}_{m,m0}$ into the $exp (-\sigma S_1)$ and the terms ${\cal
 K}^{(r)}_{m,0m}$ into the $exp (-\sigma S_2)$ separating  in this way the
 part corresponding to noninteracting Wilson loops. The remains of
 expression  (40) are responsible for the interaction, so that the explicit
 dependence of ${\cal K}^{(r)}_{m,ij},i,j\not=0$ on the distance between
 the loops enters now the answers. One is to conclude therefore that the
 latter cannot be expressed in terms of areas and string tensions, but more
 detailed knowledge of the dynamics of field strength correlators is
 required.
 When the distance between loops is much more then $T_g$ (and their average
sizes) the asymptotics of the interaction is described
usually [19] by the exchange of the glueball with lightest
mass.

\section{Discussion}

Our analysis was performed for the SU(N) gauge theory with arbitrary number
of dimensions. In the case of 1+1 gauge theory our expressions are
simplified, because in two dimensions the SU(N) theory can be
represented [20] as the abelian one choosing the axial gauge
 $$A_1^{(a)}(x_1,x_2)=0.$$ As the result, only the quadratic
 cumulant exists, and one  arrrives to the generalized area laws
 (44), (45), (47) with string tensions $\sigma^{(r)}$ in various
 representations related to the fundamental string tension $\sigma^{(f)}$ by
 the eqs. (46).

 For the U(N) theory in two dimensions  the generalized   area law for
 several Wilson loops or for the Wilson loop with selfintersections was
 obtained in [11] using the non-abelian Stokes theorem. Also on the basis of
 Migdal-Makeenko equations the Wilson loops average was calculated in 1+1
 both for U(N), $N \to \infty$ in [8], and for U(N) and SU(N) theory in
 [9].  Our expressions (44),~(45),~(47) reduced to the two-dimensional
 case coincide with the U(N) results of [8], [11] for $N \to \infty$ and
 are equivalent to the SU(N)  area laws obtained in [9]. In particular,  for
 a Wilson loop with selfintersection in the geometry of fig.$2^b$ the
 formula of [8]
  $$ <Sp(W(C_1)W(C_2))> \to N exp (-\sigma^{(f)}( S_1 -
 S_2)) (1-2\sigma^{(f)} S_2)$$ $$ N \to \infty~,~~\sigma^{(f)} \sim g^2 N~,
 $$ can be easily obtained from eq.(47).

 Let us summarize the results. We have expressed the  average of two
 Wilson loops in terms of irreducible field strength correlators. For
 large contours the generalized area law asymptotics is
 established, so that the average of two Wilson loops (in
 fundamental representation) is related to the averages of a single
 Wilson loop in the representations corresponding to the geometry
 of the contours.

   The physical interpretation of the expressions (44),(45) can be obtained if
 one considers the problem of nonperturbative field distribution between
 colour charges( in overall colour singlet state). The area law for a single
 Wilson loop corresponds to the formation of the string between
 $q\bar{q}$ in the meson. In our two loops case the generalized area law
 reflects the formation  of stringlike gluonic field configurations in four
 quarks system. We note that in the leading in $1/N$ approximation (when
 the interaction between colourless systems vanishes [21])
 $<W(C_1)W(C_2)>\to<W(C_1)><W(C_2)>$, so that one gets linear
 superposition of two mesonic strings.

 Our analysis  based on the cluster expansion method  can be easily
 generalized to evaluate the average of several Wilson loops or of a loop
 with more than one selfintersection. To this end one is to consider the
 quantity \be
 <(W(C_1))^{\alpha_1}_{\beta_1}...(W(C_n))^{\alpha_n}_{\beta_n}>
 \ee
 and decompose  the pairs of colour indeces in accordance with formulae
 generalizing expressions (17)
 or (18) depending on the geometry of the loops. The expression (48) can be
  rewritten  then as a sum over combinations corresponding to  all
 proper irreducible representations of colour group with the
 coefficients determined by the dimensions of these representations. With
 the assumption of the existence of a finite correlation  length the cluster
 expansion method allows to demonstrate that each term in this sum exhibits
 the area law behaviour and enables to calculate the string tensions
 entering this area law.

  The  authors are grateful to Yu.M.Makeenko for the discussion of two
 dimenional results and to Yu.A.Simonov for permanent interest and
 stimulating discussions.

   This work is supported by Russian Fundamental
 Research Foundation, grant N 93-02-14937.

 This paper was partly completed during the authors' stay at NIKHEF, and the
 support of the National Organization for Scientific Recearch (NWO) which
 made this stay possible, is acknowledged.

 \newpage \setcounter{equation}{0}
\renewcommand{\theequation}{A.\arabic{equation}}

{\bf Appendix A}\\

Here we prove that  only the linear combinations (26), (27) and (29) obey
the "scalar" clusterization condition. To this end we treat the problem as
an eigenvalue problem, looking for the
 numbers $\mu$ and $\nu$ for which \be \mu{\cal A}(i_1+i_2,
 j_2+j_1)+\nu{\cal B}(i_1+i_2,j_2+j_1) \to \ee $$\to \{ \mu {\cal
 A}(i_1,j_1)+\nu{\cal B}(i_2,j_2)\}\{\mu {\cal A}(i_1,j_1)+\nu{\cal
 B}(i_2,j_2)\} $$ and \be \mu {\cal A}'(i_1+i_2, j_1+j_2)+\nu{\cal
  B}'(i_1+i_2,j_1+j_2) \to \ee $$\to \{ \mu {\cal A}'(i_1,j_1)+\nu{\cal
 B}'(i_2,j_2)\}\{\mu {\cal A}'(i_1,j_1)+\nu{\cal B}'(i_2,j_2)\} $$ takes
 place at large intercluster distances.

 For the oppositely directed loops one gets from ($20$)
 $$<(a(i_1+i_2))^{\alpha}_{\beta}(b(j_2+j_1))^{\gamma}_{\delta}>=$$
 $$<(a(i_1))^{\alpha}_{\rho}(a(i_2))^{\rho}_{{\beta}}
 (b(j_2))^{\gamma}_{\sigma}(b(j_1))^{\sigma}_{\delta}>\to $$
  $$ \to<(a(i_1))^{\alpha}_{\rho}(b(j_1))^{\sigma}_{\delta}>\cdot
  <(a(i_2))^{\rho}_{{\beta}}(b(j_2))^{\gamma}_{\sigma}>,$$
  or, using eq.(22),
  \be
 {\cal A}(i_1+i_2, j_2+j_1)\to{\cal A}(i_1,j_1)
  {\cal A}(i_2,j_2)
  \ee
  \be
  {\cal B}(i_1+i_2,j_2+j_1)\to {\cal A}(i_1,j_1){\cal B}(i_2,j_2)+
  {\cal A}(i_2, j_2){\cal B}(i_1,j_1)+
  \ee
  $$N~{\cal B}(i_1,j_1){\cal
  B}(i_2,j_2)$$
  Substituting (A.3) and (A.4) into (A.1) we have
  $$\mu
  {\cal A}(i_1,j_1){\cal A}(i_2,j_2)+$$
   $$+\nu\{{\cal  A}(i_1,j_1){\cal
  B}(i_2,j_2)+{\cal A}(i_2,j_2){\cal B}(i_1j_1)+N{\cal B}(i_1,j_1){\cal
  B}(i_2,j_2)\}$$
  $$=\{\mu {\cal A}(i_1,j_1)+\nu{\cal B}(i_1,j_1)\}\{\mu
  {\cal A}(i_2,j_2)+\nu{\cal B}(i_2,j_2)\},$$ that leads to the system of
  equations
   \be
   \left\{  \begin{array}{l} \mu=\mu^2\\ \nu N=\nu^2\\
  \mu\nu=\nu
  \end{array}
  \right.
  \ee
  which has two non-trivial solutions
  $\mu_1=1,~~\nu_1 =N$ and $\mu_2=1,~~\nu_2=0$; first solution corresponds
  to the eigenfunction (26) and the second solution corresponds to the
  eigenfunction (27) (the latter is evident, if the identity
$(\lambda_a)^{\alpha}_{{\beta}}(\lambda_a)^{\gamma}_{\delta}=\frac{1}{2}\delta^{\alpha}_{\delta}
\delta^{\gamma}_{{\beta}}-\frac{1}{2N}\delta^{\alpha}_{{\beta}}
\delta^{\gamma}_{\delta}$ is used).

Similarly, for parallel loops we write the (20') as
$$<(a'(i_1+i_2))^{\alpha}_{{\beta}}(b'(j_1+j_2))^{\gamma}_{\delta}>=$$
$$<(a'(i_1))^{\alpha}_{\rho}(a'(i_2))_{{\beta}}^{\rho}(b'(j_1))^{\gamma}_{\sigma} (b'(j_2))^{\sigma}_{\delta}>\to$$
$$\to <(a'(i_1))^{\alpha}_{\rho}(b'(j_1))^{\gamma}_{\sigma}>\cdot
<(a'(i_2))^{\rho}_{{\beta}}(b'(j_2))^{\sigma}_{\delta}>,$$
or, using eq. (23),
  \be
 {\cal A}'(i_1+i_2, j_1+j_2)\to{\cal A}'(i_1,j_1)
  {\cal A}'(i_2,j_2) +
    {\cal B}'(i_1,j_1){\cal B}'(i_2,j_2)
    \ee
    \be
    {\cal B}'(i_1+i_2, j_1+j_2)\to {\cal A}'(i_1,j_1)
  {\cal B}'(i_2,j_2) +
    {\cal A}'(i_2,j_2){\cal B}'(i_1,j_1)
    \ee
    Substituting (A.5) and (A.6) into (A.2) we have
  $$    \mu\{{\cal A}'(i_1,j_1){\cal A}'(i_2,
  j_2)+{\cal B}'(i_1,j_1){\cal B}'(i_2,j_2)\}+$$
  $$+\nu\{{\cal A}'(i_1,j_1) {\cal B}'(i_2,j_2)+{\cal A}'(i_2,j_2){\cal
  B}'(i_1,j_1)\}=$$ $$ =\{   \mu {\cal A}'(i_1,j_1)+\nu{\cal B}'(i_1,
  j_1)\}\{+ \mu {\cal A}'(i_2,j_2)+\nu'{\cal B}(i_2 ,j_2)\},$$ that leads to
  the system of equations
  \be
   \left\{  \begin{array}{l} \mu=\mu^2\\ \mu
  =\nu^2\\ \nu=\mu\nu \end{array} \right.
  \ee
   with non-trivial solutions
  $\mu_1=1,\nu_1=+1$ and $\mu_2=1,\nu_2=-1$, corresponding to the
  eigenfunctions (29).

  \newpage

   \setcounter{equation}{0}
\renewcommand{\theequation}{B.\arabic{equation}}

    {\bf Appendix B}\\

  In the  case when only the second order cumulants contribute to the string
  tensions, the expressions (44), (45) and (47) are simplified
  considerably, because in this case all the cumulants $K_2^{(r)}$
  can  be expressed in terms of a quadratic correlator of the fields $\Phi$.
  As a result all string tensions can be related to the
  fundamental one. Introducing convenient notations
   \be

\delta_{mn}\Phi(z_1^{(i)},z_2^{(j)})=<A^{(m)}(z_1^{(i)})A^{(n)}(z_2^{(j)})>,~~i,j=1,2,
  \ee
  where $m,n$ are the colour octet indeces, we note first that one gets from
    (22),(23) \be {\cal A}(2,0)={\cal
  A}'(2,0)=\frac{N^2-1}{2N}\Phi(x_1^{(1)},x_2^{(1)}) \ee
  \be {\cal
  A}(0,2)={\cal A}'(0,2)=\frac{N^2-1}{2N}\Phi(x_1^{(2)},x_2^{(2)}) \ee
   \be
  {\cal B}(2,0)={\cal B}(0,2)={\cal B}'(2,0)={\cal B}'(0,2)=0 \ee \be {\cal
  A}(1,1)={\cal A}'(1,1)=-\frac{1}{2N}\Phi(x_1^{(1)},x_1^{(2)}) \ee \be
  {\cal B}(1,1)={\cal B}'(1,1)=\frac{1}{2}\Phi(x_1^{(1)},x_1^{(2)}) \ee

  Therefore one can see from (35),(39) that all the ${\cal K}_2^{(r)}$ are
  related to  the integrals
\be
 \Phi_{ij}=\frac{1}{2}\int_{S_i}\int_{S_j}
 d\sigma{(z_1}^{(i)})d\sigma(z_2^{(j)})\Phi(z_1^{(i)},z_2^{(j)})
 \ee
 in the following way
   \be
  \frac{1}{2}{\cal K}_2^{(r)}(2,0)= \frac{N^2-1}{2N}\Phi_{11},
  \ee
  \be
  \frac{1}{2}{\cal K}_2^{(r)}(0,2)= \frac{N^2-1}{2N}\Phi_{22},
  \ee
  while
  \be
  \frac{1}{2}{\cal K}_2^{(0)}(1,1)= \frac{N^2-1}{N}\Phi_{12},
  \ee
  \be
  \frac{1}{2}{\cal K}_2^{(adj)}(1,1)=
  \frac{1}{N}\Phi_{12},
   \ee
   \be
  \frac{1}{2}{\cal K}_2^{(s,a)}(1,1)=
  \pm \frac{N\mp 1}{N}\Phi_{12}.
   \ee
    For large contours embedded into the same plane we have
    \be
    \Phi_{11}=\tilde{\sigma}S_1,~~ \Phi_{22}=\tilde{\sigma}S_2,~~
 \Phi_{12}=\mp\tilde{\sigma}S_2,~~
 S_1>S_2,
 \ee
 where $-(+)$ stands for the  opposite (parallel) directions in the loops,
 and $$\tilde{\sigma}=\frac{2N}{N^2-1}\sigma^{(f)}.$$

 Substituting (B.8)-(B.13) into the eq. (43) we obtain the relations
 (46) between string  tensions  in different representations.

  \newpage

  {\bf Figure captions}\\

  Fig.1. Geometries of Wilson loops under consideration.\\

  Fig.2. Geometries of Wilson loop with one selfintersection.\\

  Fig.3. Transformation of the initial contour a) to the deformed one b) for
    the application of non-abelian Stokes theorem in the
    Fock--Schwinger gauge (4) for the case of a single Wilson
    loop.\\

    Fig.4. Deformed contours for the case of two Wilson loops
    (16), with angular correlations taken into account.
    \end{document}